\documentclass[12pt,aps,prb,preprint]{revtex4}

\usepackage{amsmath}    
\usepackage{amsfonts}
\usepackage{amssymb}
\usepackage{graphicx}
\usepackage{times}
\usepackage{psfrag}
\usepackage{epsfig}
\usepackage{multirow}

\begin{document}

\title{Protein structure prediction by an iterative search method}

\author{Ivan C. Rankenburg}
\author{Veit Elser}


\begin{abstract}
We demonstrate a new algorithm for finding protein conformations that minimize a non-bonded energy function.  The new algorithm, called the difference map, seeks to find an atomic configuration that is simultaneously in two constraint spaces.  The first constraint space is the space of atomic configurations that have a valid peptide geometry, while the second is the space of configurations that have a non-bonded energy below a given target.  These two constraint spaces are used to define a deterministic dynamical system, whose fixed points produce atomic configurations in the intersection of the two constraint spaces.  The rate at which the difference map produces low energy protein conformations is compared with that of a contemporary search algorithm, parallel tempering.  The results indicate the difference map finds low energy protein conformations at a significantly higher rate then parallel tempering.
\end{abstract}
\maketitle
\section{Introduction}

Since the three dimensional structure of a protein largely determines its function, there is tremendous incentive to determine the native structure of proteins.  For proteins that form high quality crystals, x-ray crystallographic methods have long enabled researchers to determine the protein's structure.  For proteins that fail to form high quality crystals, NMR spectroscopy is a time consuming and expensive alternative.  Currently, there are many proteins with known sequences and unknown structure.  As an alternative to x-ray crystallography and NMR spectroscopy, a reliable structure prediction method would be a tremendous asset to biological research.  The various structure prediction methods are compared at the semi-annual CASP experiment.\cite{CASP6}  In the past CASP experiments, the most successful structure prediction method has been homology modeling.  In recent years though, \textit{ab initio} methods have started to become competitive.

The method of homology modeling is based on the observation that when two proteins have similar amino acid sequences, they also usually have similar structural properties.\cite{Stuart2000}  Using this method, a protein's structure is determined first by comparing its amino acid sequence against other determined structures in the Protein Data Bank, and finding similar sequences.\cite{ZhangSkolnick2005}  For example, if a particular subsequence of amino acids almost always forms an alpha helix, then if found in the undetermined protein's sequence, the structure of this sub-sequence can be safely guessed.  In this way, the structure is piece-wise determined, and subsequently assembled. \cite{BakerSali2001}  This technique relies heavily on the availability of similar template sequences whose structures have been determined.  For large classes of proteins, such as membrane proteins, there is a dearth of templates for comparison.  For such proteins, homology modeling currently offers little promise.

With \textit{ab initio} structure prediction, the protein is modeled as a collection of atoms\cite{Wenzel2005, Liwo2006}, or united atoms \cite{irback2000, Scheraga2005}, and the native structure is assumed to be the global minimum of an appropriate energy function.\cite{Anfinsen1973}  Because the actual energy function navigated by physical proteins is difficult to  calculate precisely, there are numerous approximations in use.  Finding the global minimum of the energy function is itself a very challenging endeavor,\cite{Onuchic1997} and many different methods have also been developed for this.  All of the modern energy minimization algorithms require great computational resources, and \textit{ab initio} methods have been limited to small proteins (approximately fifty amino acids).

In this paper we consider a very simple energy function for the non-bonded interactions (explained in appendix A) and propose a new method for finding its global minimum.  The proposed method, the \textit{difference map} (DM)\cite{ElserPNAS2007}, has previously been shown to be successful at finding low energy states in an off-lattice HP model of proteins.\cite{Elser2006b, Stillinger1993}  The DM operates in a very different manner than previous search algorithms used to minimize the conformational energy.   Most energy minimization methods are based on a Monte Carlo exploration of the protein conformation's energy landscape.  For these methods, the ``iterate'' is an evolving protein conformation.  Contrasting this, the DM ``iterate'' is not a protein conformation, but an atomic configuration.  Since a polypeptide has on average about three degrees of freedom per amino acid, and an atomic configuration has three degrees of freedom per atom, the iterate of the DM searches a much larger space than that explored by Monte Carlo methods.  Searching this larger space is not necessarily a liability: deep local minima in the energy landscape that would trap a Monte Carlo iterate are easily escaped by the DM, since the DM can evolve the iterate in directions not accessible to the Monte Carlo iterate.

The DM overcomes three fundamental deficiencies of all Monte Carlo based search techniques.  First, in all Monte Carlo based searches, an entire folding pathway must be simulated in order to find the lowest energy conformation.  The DM overcomes this by immediately searching for the lowest energy state, without regard to the folding pathway.  Second, Monte Carlo search methods have a tendency of getting stuck in deep local minima of the energy landscape.  There have been many modifications to the basic method to overcome this problem\cite{Schug2005, Bachmann2005}.  However, though lessened, the problem remains to a degree.  Contrasting this, the DM escapes even deep local minima in the energy landscape, and spends very little time exploring them.  And finally, Monte Carlo search methods update the iterate by local modifications to the protein conformation, thus limiting the rate by which the protein conformation can evolve.  The DM typically makes large modifications to the configuration in each iteration.  Eventually, when the DM encounters a true fixed point, a low energy conformation has been found.

We have applied the DM algorithm to an all-atom protein model (sidechain hydrogens have been omitted, though backbone hydrogens are included for the purpose of hydrogen bonding).  The performance of the DM is compared to that of a popular Monte Carlo method, parallel tempering (PT).  To make the comparison meaningful, the two algorithms are each run on the same computer, running the same amount of time.  The atomic potential used is as simple as possible, involving only hydrophobic-hydrophilic interactions, hydrogen bonds, and steric repulsion.  Though simple, this potential is able to correctly reproduce the general structure of the native fold of the staphylococcus aurelius A protein (B domain (10-55) ).  In this paper we will refer to this protein as ``protein A''.

\section{Theory}

\subsection{Constraints and projections}

The difference map (DM) is an iterative algorithm where the iterate (an atomic configuration) is evolved by means of \textit{projections} onto two constraint spaces.  The first constraint space is the space of atomic configurations that have a valid peptide geometry.  A member of this constraint space has all bond lengths, bond angles, and left handed versus right handed orientations correct (bond lengths and angles are taken from Engh 1991\cite{enghandhuber}).  This is the space of the rotamer configurations.  Most contemporary Monte Carlo searches have the folding protein always a member of this constraint space; the energy landscape is usually viewed as an energy function on this space.  The second constraint space used by the DM is the space of atomic configurations whose non-bonded energy is less than a predefined target energy.  When freed of the peptide geometry constraint, it is easy to find a member of this constraint space.  It is clear that when an atomic configuration is found that is a member of both constraint spaces simultaneously, the problem has been solved.  In this case, an atomic configuration that has both a valid peptide geometry, and a sufficiently low energy, has been found.  The two constraint spaces are described in detail in Appendix A.

We represent an atomic configuration by $\vec{\textbf{R}} = \{ \vec{\textbf{r}}_1 , \: \vec{\textbf{r}}_2 ,  \: \dots  \}$, where $\vec{\textbf{r}}_i$ is the 3D coordinate of atom $i$.  For both constraints, the projection to that constraint space is defined as the closest atomic configuration that satisfies the constraint.  In this paper, $P_G \left[ \vec{\textbf{R}} \right]$ denotes the projection to the peptide geometry constraint, while $P_E \left[ \vec{\textbf{R}} \right]$ denotes the projection to the energy constraint.

For the geometry constraint, the projection is accomplished by minimizing a penalty function (defined in Appendix A) via an adaptive step-size steepest descent algorithm.  This projection performs a minimal modification to the atomic configuration that yields a member of the geometry constraint space.  The result of this projection has a valid peptide geometry, but non-bonded atoms are allowed to overlap, and in general there is no bias toward a low energy atomic configuration.

To compute the projection to the energy constraint, the energy function defined in Appendix A is minimized until the non-bonded energy is below a predefined target energy.  Though the result of this projection is a low energy atomic configuration, the configuration in general does not have a valid peptide geometry.  For a typical member of this constraint space, bond and angle constraints are usually not satisfied.  While computing this projection, the protein behaves as a liquid of independent atoms, rather than as a linked chain of amino acids.

\subsection{Difference map algorithm}

As a simple pedagogical step toward understanding the DM algorithm, first consider the following alternative algorithm, called \textit{alternating projections} (AP):
\[
 \vec{\textbf{R}}_{n+1}= P_G \left[ P_E \left[\vec{\textbf{R}}_n \right]\right] 
\]
For AP, the iterate is projected to the energy constraint, followed by a projection to the geometry constraint.  With the projections in this order, the iterate is perpetually a member of the geometry constraint space.  This algorithm greedily minimizes the distance between the two constraint spaces, and quickly evolves toward a fixed point, where the distance between the two constraint spaces has a local minimum.

\begin{figure}[h]
\centerline{
\epsfig{file=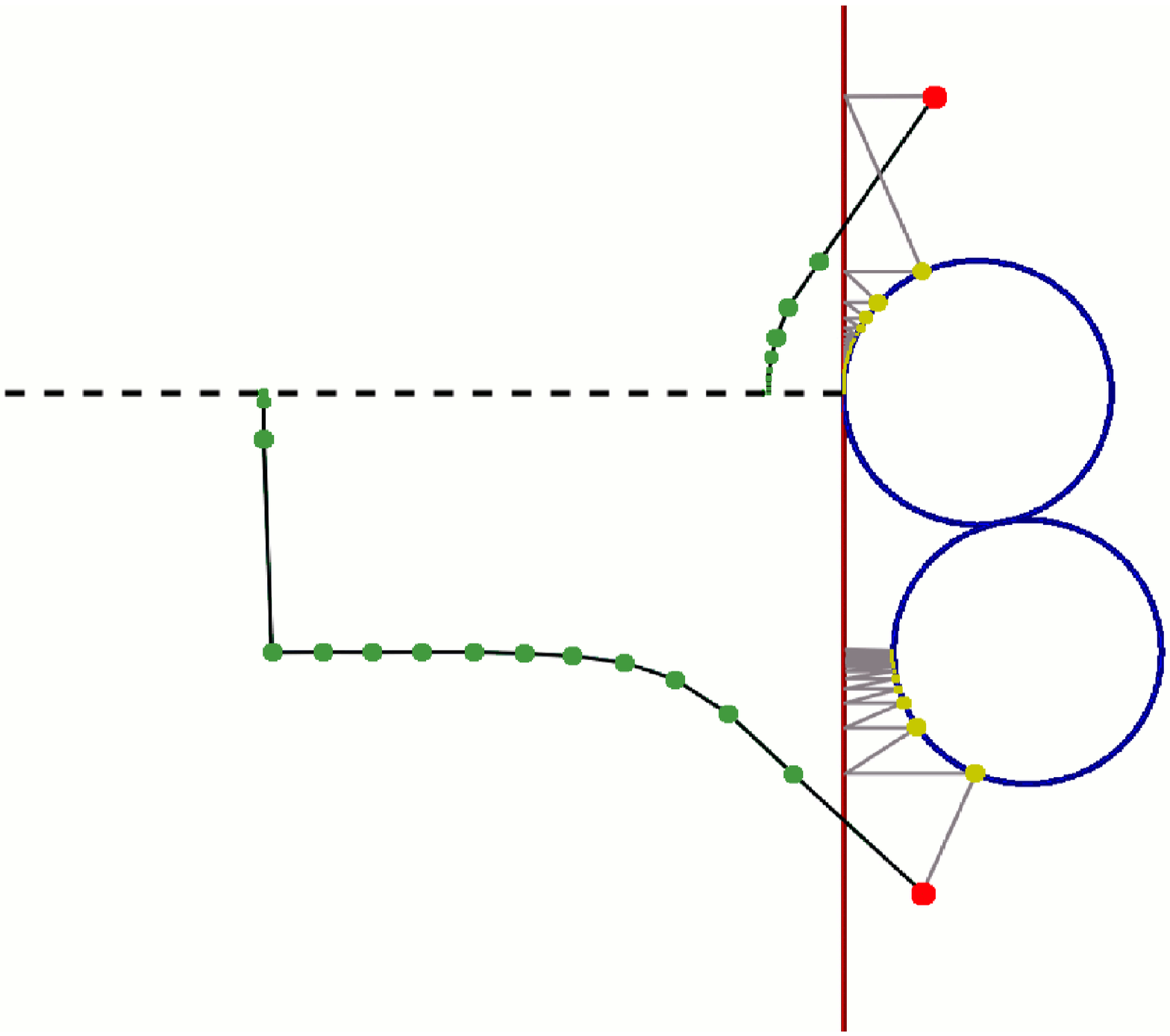, scale=0.5}
}
\caption{ A 2D example problem contrasts the search dynamics of the DM and alternating projections (AP).  The two constraint spaces are shown as red (vertical line), and blue (circles).  Two initial points (red dots) are iterated via the DM (green dots, black line) and AP (gold dots, gray line).  The dashed line is the set of fixed points of the DM.  Every fixed point of the DM is associated with the unique intersection of the constraint spaces.  For the top initial point, both search algorithms find the solution.  For the bottom initial point, AP stagnates at a near intersection of the constraint spaces, while the DM is repelled by this near intersection, and eventually finds the actual intersection.  The iterate of AP is always a member of the blue constraint space.}
\label{fig:2D_DM_vs_AP}
\end{figure}

To contrast AP and the DM, they are both applied to a 2D example problem in figure \ref{fig:2D_DM_vs_AP}.  In this example, the two constraint spaces are the red and blue curves, DM iteration is shown as green dots, and AP iteration is shown as the gold dots.  If the initial iterate is close to an actual intersection of the constraint spaces (top red dot in figure \ref{fig:2D_DM_vs_AP}), then AP will converge to the intersection.  However, AP is prone to stagnating at places where the distance between the constraint spaces is locally minimized (bottom trajectory in figure \ref{fig:2D_DM_vs_AP}).  Finally, note that the iterate of AP is always a member of the blue constraint space.

The DM was developed to remedy the stagnation problem of the AP algorithm.\cite{Elser2003a, Fienup1982}  The DM iterate is updated by $\vec{\textbf{R}}_{n+1}=\vec{\textbf{R}}_n+ \vec{\textbf{d}}$, where
\[ \vec{\textbf{d}}=
P_E \left[2 P_G [\vec{\textbf{R}}_n]- \vec{\textbf{R}}_n \right]-
P_G \left[ \vec{\textbf{R}}_n \right]  
  \: .
\]
Clearly a fixed point has been found when $ \vec{\textbf{d}}= \vec{\textbf{0}}$.  If $\vec{\textbf{R}}^*$ is a fixed point of the DM, the corresponding solution ($\vec{\textbf{R}}_{\mathrm{sol}}$) is given by 
\[
\vec{\textbf{R}}_{\mathrm{sol}}=P_E \left[2 P_G [\vec{\textbf{R}}^*]- \vec{\textbf{R}}^* \right]=P_G \left[ \vec{\textbf{R}}^* \right] \:
 .\]
Since $\vec{\textbf{R}}_{\mathrm{sol}}$ is simultaneously equal to a projection to each constraint space, it both has the correct peptide geometry, and a sufficiently low energy.  When the iterate is sufficiently near a fixed point, the attractive property of the DM leads to monotonic convergence to the fixed point.\cite{Elser2003a}

\begin{figure}[h]
\centerline{
\epsfig{file=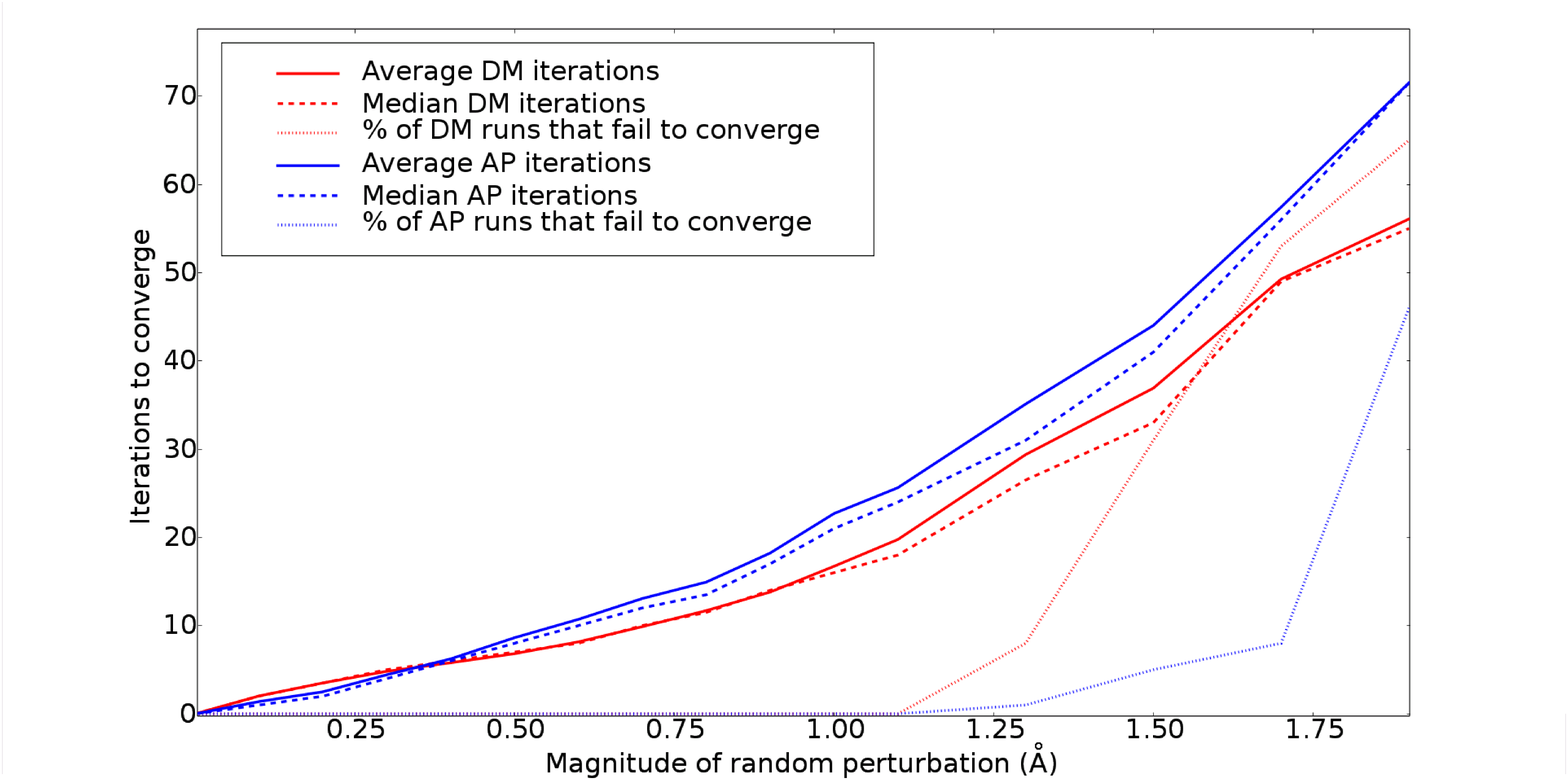, scale=0.3}
}
\caption{ The convergence rates of the DM and AP are compared.  If the atomic configuration is within 1 \AA{} RMSD of the native fold, both search algorithms always converge upon the native fold, on average within 30 iterations.  Above 1.25 \AA{} RMSD, both algorithms occasionally fail to recognize the nearby intersection of the constraint spaces.}
\label{fig:conv_rate}
\end{figure}

The extent to which the native conformation of protein A is an attractive fixed point of the DM is shown in  figure \ref{fig:conv_rate}.  Here, the initial iterate of the DM was chosen by adding random vectors of constant magnitude to protein A's atomic coordinates.  The DM then evolved the iterate, and terminated when the iterate converged upon a fixed point.  This perturbation followed by DM iteration was tested 100 times, for many different magnitudes of the perturbation.  The average number of iterations before a fixed point was found is displayed.  Given the same initial iterate, the convergence rate of the AP map was tested in the same way, and is also shown.

Though prone to stagnation, AP is a useful algorithm for finding the nearest local minimum of the conventional energy landscape.  This energy refinement is done by first projecting the atomic configuration to the geometry constraint (yielding a valid protein conformation) and evaluating the atomic configuration's non-bonded energy.  Next, the atomic configuration is projected to the energy constraint with a projection target energy only slightly lower than the current energy (this moves the iterate a small step in the downhill gradient direction).  Finally, the atomic configuration is again projected to the geometry constraint.  These alternating projections, with the target energy continually being lowered, quickly lowers the energy of the protein conformation, and eventually finds a fixed point at a nearby local minimum in the energy landscape.  These are the same local minima that could potentially trap a Monte Carlo search iterate.

To generate low energy protein conformations, the DM searched for seven days on four parallel processors, each 3 GHz.  Each processor operated independently of the others, and every search began with a configuration of random atom positions.  The initial atom coordinates where chosen from inside a box with a uniform probability distribution.

Every three hundred DM iterations, the current iterate was refined via AP until a nearby fixed point of AP was found.  The fixed point was the nearest local minimum of the conformational energy landscape, and represented the best estimate for an atomic configuration satisfying both the geometry constraint and the energy constraint.  The energy and RMSD (all atoms) of these estimates are plotted in figure \ref{fig:EVRSMD} (green dots).

After refining via AP, if the energy of an estimate was below the target energy of the energy constraint, the target energy was lowered to this new lower energy.  The iterative DM search was then restarted with a new random initial atomic configuration.  On the other hand, if after refining the energy of the estimate was above the cutoff energy, the DM iterate was replaced with the refined estimate, and DM iterations continued.

\subsection{Parallel tempering algorithm}

The difference map was compared to one of the state-of-the-art minimization algorithms, parallel tempering (PT).\cite{Swendsen1986, Swendsen2005}  PT has had significant success in folding small proteins.\cite{Earl2005, Schug2005}  The method is a modified Monte Carlo search.  For each search, there are several clones of the same initial atomic configuration.  Each clone is evolved via Monte Carlo steps at a different temperature.  At every iteration there is a probability of a swap between any two clones (a swap consists in switching their temperatures).  The probability is a function of the clones' current energies and temperatures.  Additionally, after a large number of Monte Carlo iterations, the atomic configuration of the lowest energy clone replaces the atomic configuration of the hottest clone.

The Monte Carlo step is computed by adding to each clone's atomic coordinates, random vectors of a given magnitude.  After this perturbation, the atomic configuration is projected to the geometry constraint space.  The result of these two operations is a slightly perturbed atomic configuration that has a valid peptide geometry.  After the perturbation and projection, the energy of the new protein conformation is calculated, and the probability of accepting or rejecting the test step is calculated.  The magnitude of the random perturbation is adjusted for each temperature to maintain a step acceptance rate of 50\%.

Exactly the same computational resources were applied to the PT algorithm.  We used four random initial configurations (one on each processor).  The initial atomic coordinates were generated by first choosing atom positions from inside a box with a uniform probability distribution, and then projecting the atomic configuration to the geometry constraint.  For each of the four simulations, we used fifteen clones, whose temperatures ranged from $2.92$ to $0.01$ (in the energy scale described in Appendix A).  A temperature of $2.92$ was hot enough that the clone with this temperature quickly explored the energy landscape, and spent very little time in any one energy minimum.  On the other hand, the clone with a temperature of $0.01$ was essentially frozen: its energy was low, and fluctuated only very little.  Each of the four simulations made consistent progress toward lower energies.

With our PT code, we averaged 1.7 seconds per iteration, for a single processor, with fifteen clones.  This is close to previously published iteration rates.  In their 2005 paper,\cite{Schug2005} Schug \textit{et. al.} averaged about one million iterations per fifteen clones using fifteen processors in one day, for a protein with five amino acids.  This corresponds to about ten seconds per iteration, for a single computer, with fifteen clones, iterating a protein with forty amino acids.  The fact that our PT iterations are faster is due to our comparatively simple potential.

\section{Results}

\begin{figure}[h]
\centerline{
\epsfig{file=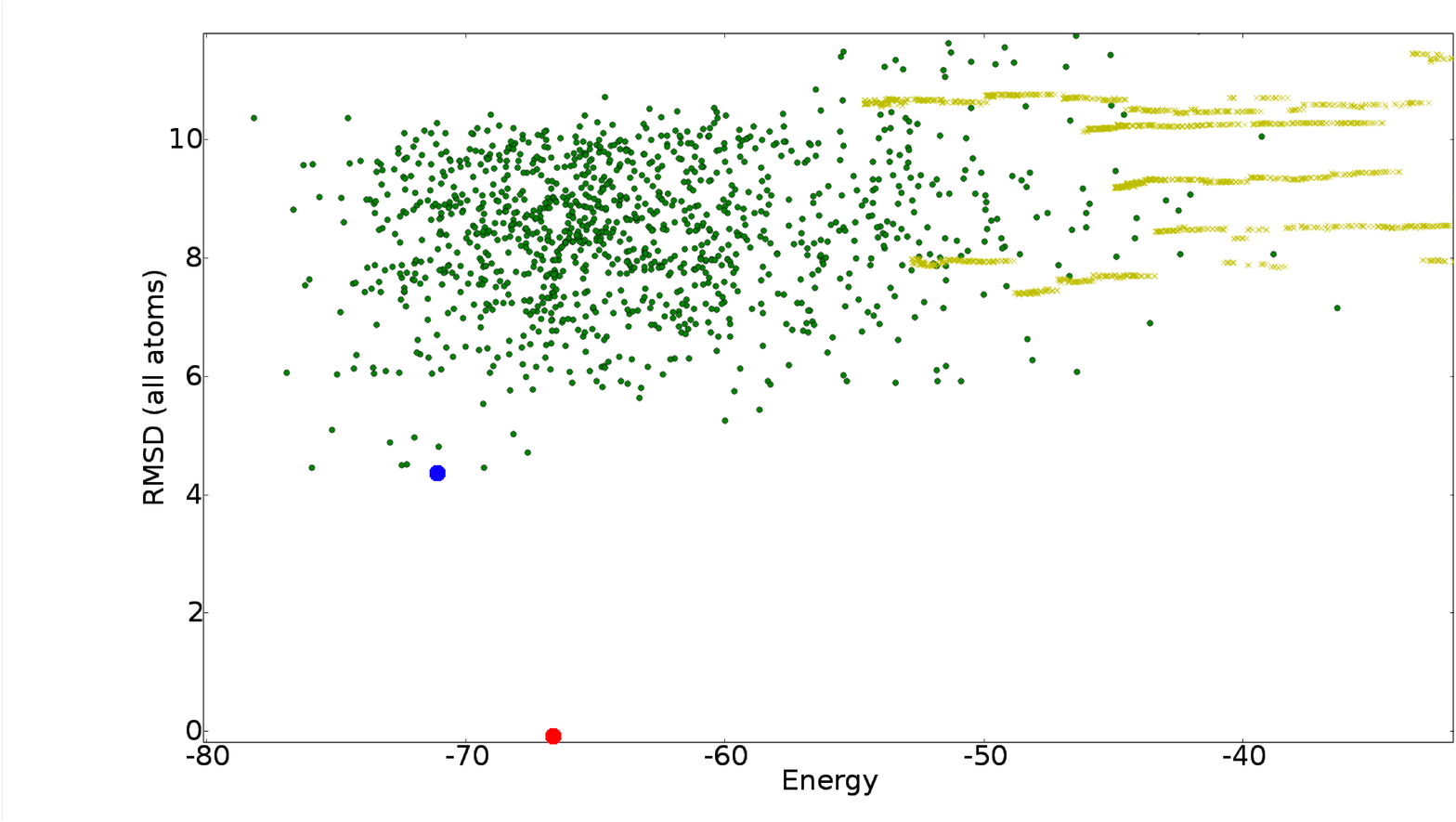, scale=0.3}
}
\caption{ Here RMSD (all atom) versus energy is plotted for the results from both search algorithms.  The green dots are the output configurations found by the DM, while the gold crosses are those found by PT.  Clearly the PT simulations are still progressing toward lower energy.  Both methods ran one week.  The blue dot is the most native-like fold discovered, and the protein conformation is shown in figure \ref{fig:bestfold}.  The red dot is the native fold, also shown in figure \ref{fig:bestfold}.}
\label{fig:EVRSMD}
\end{figure}

After one week of searching, both algorithms found many low energy atomic configurations of protein A.  The RMSD (all atom) from the native fold versus the energy of these folded proteins is shown in figure \ref{fig:EVRSMD}.  In this figure, the low energy conformations discovered by the DM are shown as green dots, and the conformations explored by PT are shown as yellow crosses.  As can be seen from the figure, the PT simulations are still progressing towards lower energy conformations.  Previous studies suggest the PT simulations will find the global minimum of the energy landscape when given enough time.

\begin{figure}[h]
\centerline{
\epsfig{file=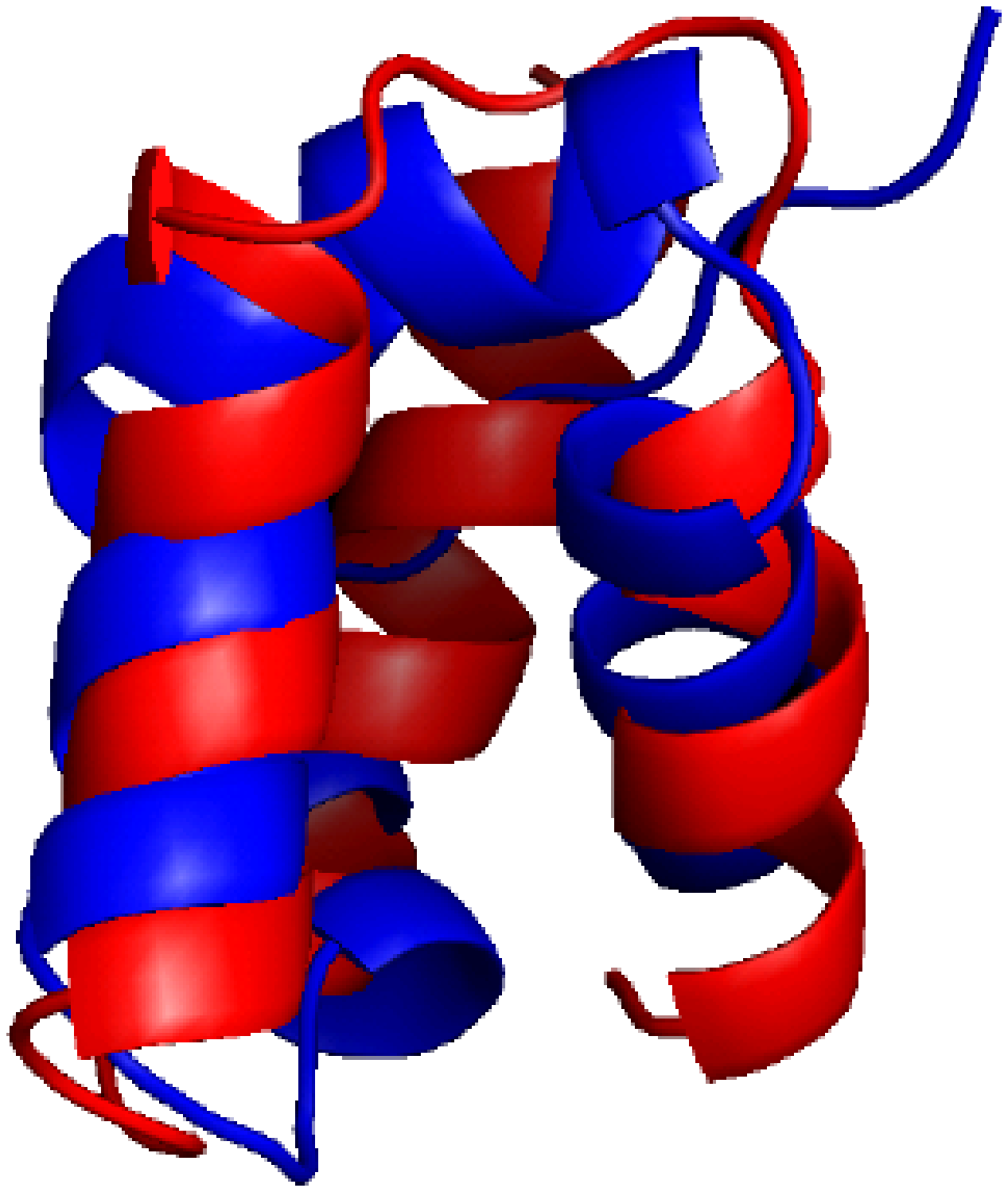, scale=0.7}
}
\caption{ The most native-like protein conformation (blue) found by the DM is shown compared to the native fold (red) of protein A.  This protein conformation was found during one week of computation.  It has an RMSD (all atom) of 4.4\AA{}, and an energy of -71.1.  The native fold has an energy of -66.6.}
\label{fig:bestfold}
\end{figure}

The lowest energy conformation the DM found had little resemblance to the native fold.  The conformation with the lowest RMSD (4.4\AA{}) found by the DM had an energy of -71.1, compared to the energy -66.6 of the native fold.  Both protein conformations are shown in figure \ref{fig:bestfold}.  We do not claim that the lowest energy protein conformation found by the DM is the lowest possible, only that given the same amount of time and resources, the DM finds many more low energy states than PT, as is evident from figure \ref{fig:EVRSMD}.  The existence of folds with energy lower than the native fold points out a deficiency of our minimalistic potential.  Here we are demonstrating the effectiveness of a new search algorithm toward the purpose of finding low energy states, rather than validating a candidate potential.

We tested several other proteins, and a range of potential parameters, and found the DM almost always finds a lower energy state than the native conformation.  This is shown in table \ref{tab:DM performance}.  To adjust the energy function, the relative scale of the hydrophobic energy to the hydrogen bonding energy was varied.  In table \ref{tab:DM performance}, the parameter $F_{HB}$ represents the fraction of the total energy due to hydrogen bonding in the native fold, and was adjusted by changing the prefactor of the hydrophobic term in the energy function.

\begin{table}[]
\caption{The performance of the DM on seven proteins, and for a range of energy parameters.  The fraction of the native conformation's energy due to hydrogen bonding is $F_{HB}$.  This is adjusted by varying the prefactor of the hydrophobic term in the energy function.  For every choice of $F_{HB}$, the native conformation's energy is given along with the best fold discovered by the DM.}
\begin{tabular}{l|@{\quad}c@{\quad}c@{\quad}c@{\quad}c}
\hline
   PDB code (length)  & $F_{HB}$       	& native	&  lowest found & search time (3GHz)\\ \hline
\multirow{3}{*}{1bba (30)}   & 0.86              & -33.2         & -34.8   &  45 hours\\
          		& 0.69                   & -50.4         & -56.0   & 9 hours \\
        		& 0.5                    & -56.7         & -60.3   & 35 hours \\ 
 \hline
\multirow{3}{*}{1enh (53)}   & 0.85              & -71.4         & -80.0   & 45 hours \\
                        & 0.62                   & -105.2        & -112.5  & 9 hours  \\
                        & 0.38                   & -125.5        & -126.6  & 35 hours \\ 
\hline
\multirow{3}{*}{1gab (45)}   & 0.89              & -61.2         & -66.7   & 45 hours \\
         	        & 0.72                   & -88.6         & -89.7   & 9 hours \\
         	        & 0.53                   & -101.0        & -102.9  & 35 hours  \\ 
 \hline
\multirow{3}{*}{1gjs (45)}   & 0.89              & -64.1         & -66.3   & 45 hours \\
         		& 0.75                   & -92.1         & -91.2   & 9 hours \\
         		& 0.59                   & -106.4        & -109.2  & 35 hours \\ 
 \hline
\multirow{3}{*}{1guu (44)}   & 0.87              & -57.8         & -65.1   & 45 hours  \\
          		& 0.70                   & -85.3         & -86.4   & 9 hours  \\
          		& 0.48                   & -102.9        & -103.7  & 35 hours   \\
 \hline
\multirow{3}{*}{1vii (35)}   & 0.84              & -43.6         & -50.1   & 45 hours \\
                        & 0.60                   & -67.1         & -71.0   & 9 hours \\
                        & 0.33                   & -82.7         & -86.7   & 35 hours \\
 \hline
\multirow{3}{*}{1ba5 (46)}   & 0.87              & -62.6         & -65.6   & 45 hours \\
           		& 0.67                   & -95.4         & -94.0   & 9 hours \\
          		& 0.43                   & -111.0        & -116.4  & 35 hours \\ 
 \hline
\end{tabular}
\label{tab:DM performance} 
\end{table}

\section{Discussion}

As can be seen in figure \ref{fig:EVRSMD}, the PT data points form an almost continuous trajectory.  There were four different PT simulations, and there can be seen four yellow streaks, each occasionally broken by a discontinuity.  By being constrainted to move in the usual energy landscape, PT is forced to make small modifications to the protein conformation.  Because of this, the PT simulations in effect reconstruct the folding pathway.  If the goal is to find the lowest energy conformation, however, then simulating the entire folding pathway is unnecessary.  While the PT algorithm was simulating the folding pathway of the protein, the DM was searching for low energy conformations directly, with no regard for the physical pathway.  This accounts, to a large extent, for the superior performance of the DM algorithm.

The most native-like protein conformation produced by the DM is shown in figure \ref{fig:bestfold}.  This conformation, like the native fold, has three helices.  The DM also found many lower energy folds --- a defect of our minimalistic potential.  Our computationally simple potential nevertheless enables us to show the effectiveness of the DM search algorithm, as compared to PT.   Several choices of potential parameters were explored in the process of trying to find a potential with the property that the native conformation is the lowest energy fold, but this proved to be impossible for the eight proteins studied.  We believe the superior performance of the DM algorithm over PT will extend to more realistic potentials as well.

In this paper we demonstrate a new search algorithm, based not on the physical pathway of the folding process, but on the geometry of constraint spaces.  Our results show the difference map  algorithm is very efficient for finding low energy states for a given potential.  The algorithm is both easy to implement, and is easy to run in parallel.  It is our hope that this new method for finding low energy atomic configurations will facilitate the development of more precise atomic potentials, since the most important feature of a good potential is that the native fold has the lowest energy.

\bibliography{ref}

\appendix
\addcontentsline{toc}{chapter}{Appendices}
\section{}
\subsection{Geometry constraint}

The geometry constraint ensures all the bond lengths and angles are correct (data from Engh 1991\cite{enghandhuber}), all the peptide bonds are in the trans orientation, and the Ramachandran angles are not in the sterically forbidden region of the Ramachandran plot.  An atomic configuration satisfying these conditions is a valid protein conformation (rotamer).  We use the following penalty function to implement the geometry constraint:
\begin{equation}
\label{GeomEQ}
0 = \sum\limits_{i \in \text{bonds}} B_i + \sum\limits_{i \in \text{angles}} A_i+\sum\limits_{i \in \text{dets}} D_i+\sum\limits_{i \in \text{peptide bond}} \Omega_i+\sum\limits_{i \in \text{ramas}} R_i
\end{equation}
For the first term of equation~\eqref{GeomEQ}, the index $i$ runs over all the bonds in the protein.  For each bond, $B_i=p_i\left( \vec{\textbf{v}}_i \cdot \vec{\textbf{v}}_i-b_i^2 \right)^2$.  Each bond has a target length, $b_i$, a penalty weight $p_i$, and $\vec{\textbf{v}}_i$ is the vector connecting the two atoms participating in the bond.  Essentially, each $B_i$ is a measure of how correct the $i^{\text{th}}$ bond is, and $p_i$ is the relative cost of the $i^{\text{th}}$ bond deviating from correct.  For all backbone bonds (such as $\text{C}_\alpha$-C, C-N, or N-$\text{C}_\alpha$ ) $p_i$ is 4, for the bonds coming directly off the backbone (such as C-O, N-H, or $\text{C}_\alpha$-$\text{C}_\beta$ ) $p_i$ is 2, and the bonds within a sidegroup are given a penalty weight of 1.

In the second term of equation~\eqref{GeomEQ}, the index $i$ runs over all the bond angles in the protein.  The $i^{\text{th}}$ angle is defined by two vectors, $\vec{\textbf{v}}_{i,1}$ and $\vec{\textbf{v}}_{i,2}$, and $A_i=p_i\left( \vec{\textbf{v}}_{i,1} \cdot \vec{\textbf{v}}_{i,2}-a_i \right)^2$.  Every angle has a target dot product for the two vectors, $a_i$, and a penalty weight $p_i$.  Note that the magnitudes of $\vec{\textbf{v}}_{i,1}$ and $\vec{\textbf{v}}_{i,2}$ are each controlled by the bond constraint above.  Like $B_i$ above, $A_i$ is a measure of how correct the $i^{\text{th}}$ angle is, and $p_i$ is the relative cost of the $i^{\text{th}}$ angle deviating from correct.  For all backbone angles (such as $\text{C}_\alpha$-C-N, C-N-$\text{C}_\alpha$, N-$\text{C}_\alpha$-C) $p_i$ is 2, for angles involving bonds directly off the background (such as O-C-$\text{C}_\alpha$, O-C-N, H-N-$\text{C}_\alpha$, H-N-C, $\text{C}_\beta$-$\text{C}_\alpha$-C, and $\text{C}_\beta$-$\text{C}_\alpha$-N) $p_i$ is 1, and for all angles within a sidegroup $p_i$ is 0.5.

\begin{figure}[h]
\centerline{
\epsfig{file=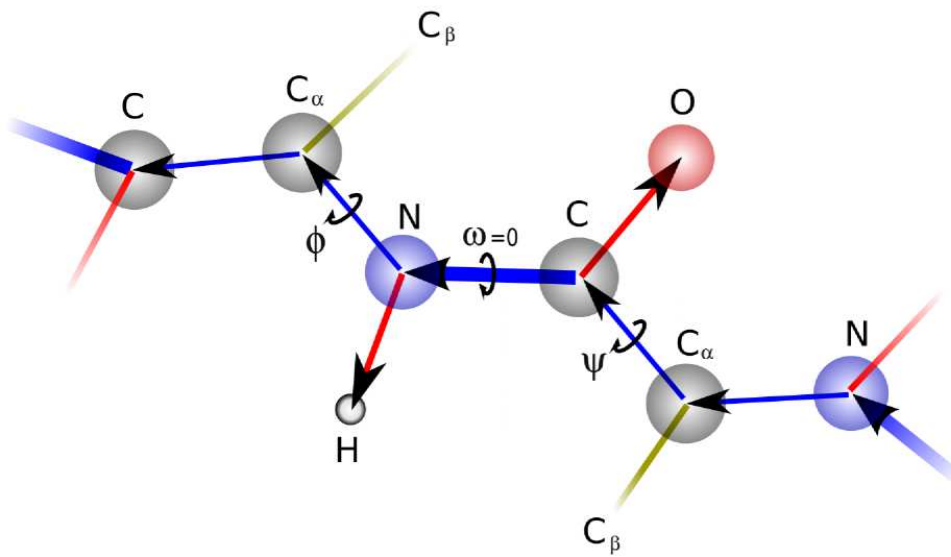, scale=0.45}
}
\caption{ A schematic of the atoms involved in the peptide bond.  The atoms H, N, C, and O all lie in a plane.  The H and O are in the trans configuration.}
\label{fig:con_atoms}
\end{figure}

For the third term of equation~\eqref{GeomEQ}, the index $i$ runs over backbone peptide bonds.  These $\Omega_i$ terms ensure that backbone hydrogens and oxygens are in the trans configuration, and that the atoms H, N, C, and O all lie in a plane.  Figure \ref{fig:con_atoms} shows the correct trans orientation of the atoms participating in a peptide bond, and will aid in visualizing the vectors described below.  To ensure the trans configuration, terms $\Omega_i=p_i \left( \vec{\textbf{v}}_{i,1} \cdot \vec{\textbf{v}}_{i,3}-d_0 \right)^2$ are included.  Here, $\vec{\textbf{v}}_{i,1}$ is the C-O vector and $\vec{\textbf{v}}_{i,3}$ is the N-H vector for the $i^{\text{th}}$ peptide bond, and $d_0$ is the correct dot product of these two vectors.  The penalty weight $p_i$ is 1.5.  To ensure the atoms H, N, C, and O all lie in a plane, terms $\Omega_i=p_i \left( \vec{\textbf{v}}_{i,1} \cdot (\vec{\textbf{v}}_{i,2} \times \vec{\textbf{v}}_{i,3})\right)^2$ are included.  Here, $\vec{\textbf{v}}_{i,1}$ and $\vec{\textbf{v}}_{i,3}$ have the same meaning as before, $\vec{\textbf{v}}_{i,2}$ is the C-N peptide bond vector, and the penalty weight $p_i$ is 0.2 .  To ensure the trans configuration of consecutive $\text{C}_\alpha$'s, and to ensure that the four atoms $\text{C}_\alpha$, C, N, and $\text{C}_\alpha$ all lie in a plane, there are identical constraints involving these four atoms.  Note that this term to ensure planarity is redundant. The relevant angle constraints would, on their own, ensure the atoms lie in a plane.  However, this additional, more direct planar constraint makes the overall projection to this constraint set converge faster.

The fourth term of equation~\eqref{GeomEQ} is for four-atom configurations where left-handed versus right-handed orientations are relevant.  For every amino acid (except glycine) the four atoms $\text{C}_\beta$, $\text{C}_\alpha$, C, and N define a parallelepiped, whose volume is constrained by $D_i= p_i \left( \vec{\textbf{v}}_{i,1} \cdot (\vec{\textbf{v}}_{i,2} \times \vec{\textbf{v}}_{i,3}) -V_0 \right) ^2$, where $\vec{\textbf{v}}_{i,1}$ is the $\text{C}_\alpha$-$\text{C}_\beta$ vector, $\vec{\textbf{v}}_{i,2}$ is the $\text{C}_\alpha$-C vector , $\vec{\textbf{v}}_{i,3}$ is the $\text{C}_\alpha$-N vector, $V_0$ is the target parallelepiped volume (note the sign of $V_0$ dictates left handed versus right handed orientations), and the penalty weight $p_i$ is 1.  Also, for sidegroups that have a left handed versus right handed preference, such as the side group of isoleucine, there is an identical constraint relating the orientation of the relevant atoms.

\begin{figure}[h]
\centerline{
\epsfig{file=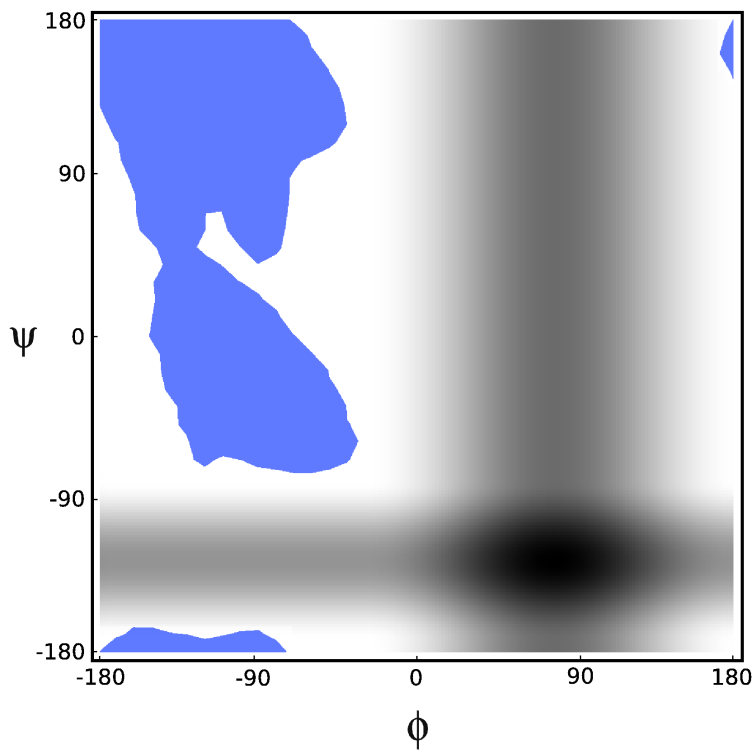, scale=.5}
}
\caption{ The natural distribution of Ramachandran angles (blue area) is shown along with a greyscale of the Ramachandran penalty used in equation~\eqref{GeomEQ}.  The distribution data is taken from Kleywegt 1996.\cite{Kleywegt1996}  98\% of all non-glycine Ramachandran angles lie in the blue shaded area.}
\label{fig:rama_dist}
\end{figure}

The last term of equation~\eqref{GeomEQ} controls the range of the Ramachandran angles $\phi$ and $\psi$.  Due to steric repulsion, there is a significant region of the Ramachandran plot that is inaccessible to proteins, shown in figure \ref{fig:rama_dist}.  Rather than calculating $\phi$, it is easier to calculate the sine and the cosine of $\phi$ by:
\[
\sin \phi=\vec{\textbf{v}}_2 \cdot (\vec{\textbf{v}}_3 \times \vec{\textbf{v}}_1) \;
\frac{\vert \vec{\textbf{v}}_2 \vert}{\vert \vec{\textbf{v}}_1 \times \vec{\textbf{v}}_2 \vert \: \vert \vec{\textbf{v}}_2 \times \vec{\textbf{v}}_3 \vert }
 \]\[
\cos \phi=
\left[ (\vec{\textbf{v}}_1 \cdot \vec{\textbf{v}}_2) \; (\vec{\textbf{v}}_2 \cdot \vec{\textbf{v}}_3) - 
(\vec{\textbf{v}}_1 \cdot \vec{\textbf{v}}_3) \; (\vec{\textbf{v}}_2 \cdot \vec{\textbf{v}}_2)  \right] \;
\frac{1}{\vert \vec{\textbf{v}}_1 \times \vec{\textbf{v}}_2 \vert \: \vert \vec{\textbf{v}}_2 \times \vec{\textbf{v}}_3 \vert }
\]
where $\vec{\textbf{v}}_1$, $\vec{\textbf{v}}_2$, and $\vec{\textbf{v}}_3$ are the three vectors defining the torsion angle $\phi$.  Specifically, $\vec{\textbf{v}}_1$ is the C-N vector, $\vec{\textbf{v}}_2$ is the N-$\text{C}_\alpha$ vector, and $\vec{\textbf{v}}_3$ is the $\text{C}_\alpha$-C vector (see figure \ref{fig:con_atoms}).  Similarly, the sine and the cosine of $\psi$ are calculated in the same way, except $\vec{\textbf{v}}_1$, $\vec{\textbf{v}}_2$, and $\vec{\textbf{v}}_3$ are the three vectors defining the torsion angle $\psi$, or $\vec{\textbf{v}}_1$ is the N-$\text{C}_\alpha$ vector, $\vec{\textbf{v}}_2$ is the $\text{C}_\alpha$-C vector, and $\vec{\textbf{v}}_3$ is the C-N vector.

In terms of $\phi$ and $\psi$, the $R_i$ in equation~\eqref{GeomEQ} is the sum of a function controlling the $\phi$ distribution and the $\psi$ distribution.  Specifically, 

\begin{eqnarray*}
R_i&=&
\left\{
 \begin{array}
{l@{\quad \text{if} \quad}l}
0 & f(\phi) < -0.26 \\
0.3  \left( f(\phi) + 0.26 \right)^2 & f(\phi) > -0.26 
 \end{array} \right. \\
&+&  \left\{
 \begin{array}
{l@{\quad \text{if} \quad}l}
0 & g(\psi) > -0.50  \\
6.0  \left( g(\psi) + 0.50 \right)^2 & g(\psi) < -0.50 
 \end{array} \right. \\
&&\\
\text{where} \quad f(\phi)&=& 0.83 \; \sin \phi + 0.17 \; \cos \phi \\
\text{and} \quad   g(\psi)&=& 0.56 \; \sin \psi + 0.44 \; \cos \psi
\:.
\end{eqnarray*}
$R_i$ is shown by the shading in figure \ref{fig:rama_dist}.  The region defined by $R_i>0$ crudely approximates the region of the Ramachandran plot inaccessible due to steric repulsion, also shown in figure \ref{fig:rama_dist}.

A member of the geometry constraint has each of the penalty functions above equal to zero.  For the projection to this constraint space, all five terms of equation~\eqref{GeomEQ} are minimized by an adaptive step-size steepest descent algorithm, and are considered sufficiently close to zero when the total penalty is less than 0.001 per amino acid.  The various energy weights $p_i$ used above were chosen such that this minimization is computed efficiently, and never frustrated.  After the minimization, all of the bonds have the proper length, all of the angles are correct, all of the peptide bonds are planar and in the trans configuration, and all of the Ramachandran angles are in the allowed region of figure \ref{fig:rama_dist}.  However, an atomic configuration satisfying the geometry constraint may have non-bonded atoms overlapping.  It is the fact that atoms are allowed to pass through each other that makes the minimization of the penalty function always successful, and never frustrated.

\subsection{Energy constraint}

The energy constraint set is defined as the set of atomic configurations with a non-bonded energy below a given target, $E_0$.  The constraint space is thus dependent on the energy target $E_0$.  An atomic configuration satisfying this condition does not necessarily have a valid peptide geometry, indeed bonded atoms may be quite separated, and the atomic configuration may not even resemble a polypeptide.  In detail, this constraint is determined as follows:
\begin{equation}
\label{EnergEQ}
E_0 > E_{NB}=\sum\limits_{i,j \in \text{atoms}} \text{VE}_{ij} +\sum\limits_{i,j \in \text{atoms}} \text{HP}_{ij} +\sum\limits_{i \in \text{H} , \: j \in \text{O}} \text{HB}_{ij}
\end{equation}
The first term of equation~\eqref{EnergEQ} prevents atoms from overlapping.  Specifically,
\[
\text{VE}_{ij}=
\left\{
 \begin{array}
{l@{\quad \text{if} \quad}l}
0 & r_{ij}>r_0 \\
\left( 1-\left( \frac{r_{ij}}{r_0} \right) ^2  \right) ^2 & r_{ij}<r_0
 \end{array} \right.
\]
where $r_{ij}$ is the distance between the $i^{\text{th}}$ and $j^{th}$ atoms, and $r_0$ is the distance at which the atoms start overlapping.  The $r_0$ used is a sum of the atomic van der Waals radii of the the $i^{\text{th}}$ and $j^{th}$ atoms.  The following radii are used: 1.57\AA{} for aliphatic sidegroup carbon, 1.41\AA{} for aromatic sidegroup carbon, 1.44\AA{} for backbone carbon, 1.34\AA{} for nitrogens, 1.20\AA{} for oxygens, 0.65\AA{} for hydrogens, and 1.57\AA{} for sulfur.  These radii are based on data from a previous study,\cite{lomize2002} and have been adapted to our $\text{VE}_{ij}$ functional form.  Note $\text{VE}_{ij}$ goes to $1$ at an atomic separation of $r_{ij}=0$, and smoothly goes to zero at $r_{ij}=r_0$.  Atoms that are bonded together must be treated specially, since they must be allowed to come closer together than non-bonded atoms.  For these pairs, $r_0$ is 40\% of the sum of constituent atomic van der Waals radii.

\begin{table}[]
\caption{\label{tab:hydro_energies} \textbf{Hydrophobicity values}  A negative number means the two atom types attract each other, positive numbers indicate repulsion.  These parameters are based on those found in a previous study\cite{lomize2002}, and have been adapted to our functional form of $\text{HP}_{ij}$.}
\begin{tabular}{lcccc}
                           & Aliphatic Carbon & \: Aromatic Carbon & \: Polar & \: Sulfur \\
Aliphatic Carbon           & -0.108                       & -0.075                     & 0.072             & -0.063 \\
Aromatic Carbon            & -0.075                       & -0.081                     & 0.093             & -0.048 \\
 Polar                     & 0.072                        & 0.093                      & 0.126             & 0.084 \\
 Sulfur                    & -0.063                       & -0.048                     & 0.084             & -0.126
\end{tabular}

\end{table}

The second term of equation~\eqref{EnergEQ} simulates the entropic interaction of water with various sidegroup atoms.  Two atoms only interact via this hydrophobic energy if they are both sidegroup atoms ($\text{C}_\beta$, $\text{C}_\gamma$, etc.), and they belong to different amino acids.  The functional form of $\text{HP}_{ij}$ is,
\[
\text{HP}_{ij}=\left\{ \begin{array}
{l@{\quad \text{if} \quad}l}
E_{ij} \left( 2\left( \frac{r_0}{r_{ij}} \right) ^2 -\left( \frac{r_0}{r_{ij}} \right) ^4 \right) & r_{ij}>r_0\\
E_{ij} & r_{ij}<r_0
 \end{array} \right.
\]
where $r_0$ is calculated the same as above, and $E_{ij}$ is the interaction energy depending on the participating atom types.  The energies used are shown in table \ref{tab:hydro_energies}.  Since hydrophobic atoms, in this model, attract each other, they tend to form a well defined oily core, while polar atoms repel every atom type, and thereby tend to inhabit the surface of the protein.

\begin{figure}[h]
\centerline{
\epsfig{file=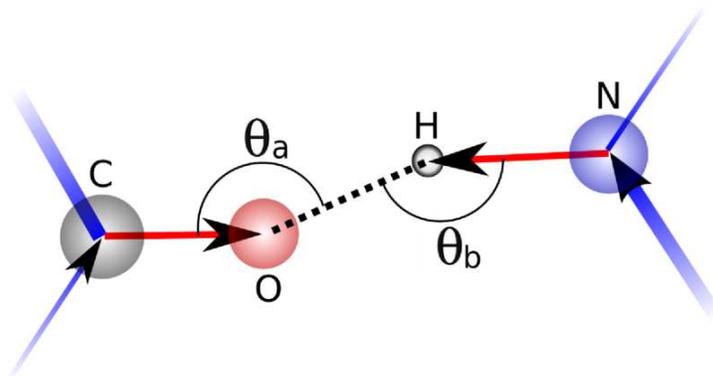, scale=0.4}
}
\caption{ The definition of $\theta_a$ and $\theta_b$ for hydrogen bonding.  The dotted line is the hydrogen bond.}
\label{fig:HB}
\end{figure}

The final energy of equation~\eqref{EnergEQ} represents hydrogen bonding.  The indices $i$ and $j$ run over all the backbone hydrogen atoms and all the backbone oxygen atoms respectively.  The functional form of $\text{HB}_{ij}$ is the product of a distance dependent function $f(r_{ij})$ and a function of the two angles $g(\theta_a , \theta_b)$ formed by the C-O-H angle $\theta_a$, and the O-H-N angle $\theta_b$ (see figure \ref{fig:HB}).  This energy function is essentially a rescaling of the one used by Irb\"ack \textit{et. al.} 2000.\cite{irback2000}  In terms of the distance function$f(r_{ij})$ and the angular function $g(\theta_a , \theta_b)$, $\text{HB}_{ij}$ is,

\begin{eqnarray*}
\text{HB}_{ij}&=&f(r_{ij}) \; g(\theta_a , \theta_b) \\
\text{where} &&\\
f(r_{ij}) &=& \left\{ \begin{array}
{l@{\quad \text{if} \quad}l}
-1.5 \left( 2\left( \frac{r_0}{r_{ij}} \right) ^2 -\left( \frac{r_0}{r_{ij}} \right) ^4 \right) & r_{ij}>r_0\\
-1.5 & r_{ij}<r_0
 \end{array} \right. \\
g(\theta_a , \theta_b)&=&\left\{ \begin{array}
{l@{\quad  \quad}l}
 \cos^2 \theta_a  \cos^2 \theta_b  & \text{if} \quad \cos\theta_a >0 \; \text{and} \; \cos \theta_b >0\\
0 & \text{otherwise}\\
 \end{array} \right.
\end{eqnarray*}
The target hydrogen bond distance, $r_0$, is 1.9\AA, and the distance between the $i^{\text{th}}$ hydrogen and the $j^{\text{th}}$ oxygen is $r_{ij}$.  Also, two backbone atoms can form hydrogen bonds only if they are separated by at least two other amino acids.

The three energies in equation~\eqref{EnergEQ}, in total, constitute the non-bonded energy of an atomic configuration.  The energy constraint space is defined as the set of atomic configurations whose non-bonded energy is less than a predefined target energy, $E_0$.  The projection to this constraint space is done by minimizing the total energy with an adaptive step-size steepest descent algorithm until the total energy is equal to $E_0$. 

Note finally that an atomic configuration whose non-bonded energy is less than $E_0$ (and therefore a member of this constraint space) does not necessarily have a valid peptide geometry.  For example, it is possible to have two bonded atoms separated by large distances, and the atomic configuration can still be a member of this constraint space.  The energy functions above treat the atomic configuration as a collection of independent atoms, rather than a linked chain.

\end{document}